# Mechanism Design for Stable Matching with Contracts in a Dynamic Manufacturing-as-a-Service (MaaS) Marketplace

Deepak Pahwa[a], Umut Dur[b], Binil Starly[a],[*]
[a] Edward. P. Fitts Department of Industrial and Systems Engineering
111 Lampe Drive, North Carolina State University, Raleigh, NC 27695
[b] Department of Economics
Nelson Hall 4108, North Carolina State University, Raleigh, NC 27695
[*]Contact author: bstarly@ncsu.edu, Tel: 919 515 1815, Fax: 919 515 5281

**Abstract**

Two-sided manufacturing-as-a-service (MaaS) marketplaces connect clients requesting manufacturing services to suppliers providing those services. Matching mechanisms i.e. allocation of clients' orders to suppliers is a key design parameter of the marketplace platform. The platform might perform an allocation to maximize its revenue or optimize for social welfare of all participants. However, individual participants might not get maximum value from their match and reject it to form matches (called blocking groups) themselves, thereby bypassing the platform. This paper considers the bipartite matching problem in MaaS marketplaces in a dynamic environment and proposes approximately stable matching solutions using mechanism design and mathematical programming approaches to limit the formation of blocking groups. Matching is based on non-strict, incomplete and interdependent preferences of participants over contracts enabling negotiations between both sides. Empirical simulations are used to test the mechanisms in a simulated 3D printing marketplace and to evaluate the impact of stability on its performance. It is found that stable matching results in small degradation in social welfare of the marketplace. However, it leads to a significantly better outcome in terms of stability of allocation. Unstable matchings introduce anarchy into marketplace with participants rejecting its allocation leading to performance poorer than stable matchings.

**Keywords:** Mechanism Design, Matching with Contracts, Manufacturing-as-a-Service Marketplaces, Dynamic Matching, Bipartite Matching, Almost Stable Matching

## 1 Introduction

Two-Sided Manufacturing-as-a-Service (MaaS) marketplace platforms connect clients (e.g. product designer) requesting manufacturing services to suppliers providing those services. In the context of quick turn-around prototyping services and low quantity production runs, the platform has the ability to provide instant quotes and have the orders placed almost immediately. Clients obtain access to larger capacities and a variety of manufacturing capabilities compared to their known traditional supplier network. On the other side, suppliers benefit through an additional source of revenue by selling their underutilized capacity. In a MaaS marketplace, the intermediary platform performs the task of order allocation and it is important that both types of participants find value in the allocation performed by the platform. If dissatisfied, the individual participants can reject the allocation and collaborate outside the platform or stop using it completely. Therefore, it is important for the platform to ensure that the order allocation or matching is stable (Gale and Shapley, 1962). A stable matching ensures that no two participants (clients / suppliers) will prefer each other over their assigned match. In other words, the participants receive their best available match given a set of suppliers and clients during a matching period. Research has shown that stable matches lead to a more successful marketplace (Roth, 2008) as the individual participants receive their fair share of value from the marketplace and do not have incentive to bypass the platform.

This work proposes matching mechanisms which enforce stability in order allocation in a MaaS marketplace ensuring that individual participants receive their fair share of value from the marketplace making it sustainable in the long term. Order allocation is done by the platform while taking into account the preferences of clients and suppliers. Clients have preferences over attributes of suppliers such as preferring to work within a supplier's geographic location and suppliers have preferences over attributes of clients such as preferring jobs with certain material or client type. Such preferences are 1) Non-strict i.e. two or more suppliers can be indifferent for a client and vice versa; 2) Incomplete i.e. a client would prefer to remain unmatched over matching with a subset of suppliers and vice versa, and 3) Cardinal i.e. the utility received from a match can be quantified. Moreover, suppliers can accept multiple orders and their preferences over orders are interdependent. For example, a small order requiring 1 hour of production time can be one of the lowest preferences of a supplier which has an available capacity of 9 hours but it becomes important when combined with two other orders that require 8 hours of combined capacity.

Negotiations can play an important role in any MaaS marketplace. Clients can approach different suppliers and negotiate with them over attributes such as price or delivery date. Therefore, a client can have varying preferences over the same supplier depending on the attributes of the contract the supplier is willing to accept. This paper considers matching with contracts framework (Hatfield and Milgrom, 2005) where participants on both sides have preferences over contracts. Additionally, the matching problem in a MaaS marketplace is dynamic in nature with stochastic arrivals of clients and suppliers. The match possibilities change with time in a dynamic setting and a myopic policy where agents are immediately matched on arrival might have short run benefits but disadvantageous in the long run. Therefore, the platform may wait, termed as "Strategic Waiting" (Liu et al., 2019) for the market to thicken before allocating orders.

This work considers the bipartite matching problem in MaaS marketplaces in a dynamic environment where matching is performed repetitively over time and develops approximately stable matching solutions. Participants have non-strict, incomplete and interdependent preferences (for suppliers) over contracts on which the matching is performed. Mechanism design and mathematical programming approaches are used to develop these solutions. The approximately stable matching solutions are compared with a socially optimal matching to quantify the degradation in system performance due to enforcing stability. The contributions of this work can be summarized as follows:

- An approximately stable matching algorithm based on mechanism design and a maximum weight approximately stable matching solution using mathematical programming formulation are proposed to achieve stability in order allocation in a MaaS marketplace. A socially optimal (but unstable) matching is also formulated to compare its performance with approximately stable solutions. The comparison quantifies the degradation in matching performance of the marketplace by enforcing stability. It also evaluates the performance of all three solutions in terms of stability. Stable matching solutions are important for the success of a marketplace as demonstrated in other two-sided markets (Roth, 2008). Unstable matches lead to participants bypassing the marketplace resulting in anarchy and inefficiency.

- The proposed solutions relaxes a key assumption considered in previous literature studying MaaS marketplaces where preferences of suppliers are specified over individual orders and are independent. In this work, suppliers are considered to be having interdependent preferences over groups of orders in which case a stable matching might not always exist. Additionally, matching is performed over contracts which enables negotiations between the participants on two sides over terms of the contracts. These considerations make the proposed solutions more viable since it takes into account realistic behavior in an actual marketplace. The proposed solutions are evaluated in a simulated MaaS platform under different assumptions on suppliers' and clients' behavior in the marketplace.

The remainder of the paper is organized as follows. Section 2 presents literature review, Sections 3 and 4 present problem description and proposed matching models respectively. Section 5 discusses the stability of proposed solutions. Section 6 describes the empirical simulation environment and Section 7 presents the results. Section 8 discusses ideas for further exploration and Section 9 draws conclusions from this work.

## 2   Literature Review

The concept of stability was first introduced in the stable marriage problem by Gale and Shapley (Gale and Shapley, 1962). It provides a stable solution in polynomial time and always finds a stable matching. It has been extended to applications such as matching doctors to medical residency programs (Roth, 1984; Manlove et al., 2017), school choice (Balinski and Sönmez, 1999; Abdulkadiroğlu and Sönmez, 2003) and others (Zhang et al., 2017; Wang et al., 2018). Refer to Roth and Sotomayor (1992) and Iwama and Miyazaki (2008) for a detailed review on the literature of stable marriage problem. Papers working on the stable marriage problem usually require the preferences of the participants on each side to be strict and complete. Gale and Sotomayor (1985) showed that a stable matching can be found in polynomial time in case of incomplete preferences whereas Irving (1994) showed that a weakly stable matching can be found in polynomial time in case of indifferent preferences. In a weakly stable matching, no two participants will strictly prefer each other over their assigned match. In case of both relaxations, indifferent and incomplete preferences, Iwama et al. (1999) show that a weakly stable matching exists for any matching instance.

In case of many to one matchings with indifferent, incomplete and interdependent preferences, as is the case with MaaS marketplaces, a weakly stable matching is not guaranteed for all matching instances. Therefore, it becomes important to find matchings which are close to stable (Manlove et al., 2017). Also, multiple stable matchings can exist for a matching instance in a MaaS marketplace and finding the stable matching with the largest size (largest number of orders accepted) or the stable matching which maximizes a utility objective (collective welfare of all participants) is NP-hard (Iwama et al., 1999). Finding these specific stable matchings is an important consideration in a MaaS marketplace. Vate (1989), Rothblum (1992), Roth et al. (1993), Abeledo and Blum (1996) and Vohra (2012) study the stable marriage problem using linear programming approaches. Stable matching is characterized as a polytope whose extreme points are stable. Since there exists multiple stable matchings for a matching instance, using mathematical programming approaches are helpful to find the stable matching which maximizes a utility objective. Matching with contracts (Hatfield and Milgrom (2005), Hatfield and Kojima (2009, 2010)) is a framework where agents on both sides have preferences over contracts rather than preferences over the agents on the other side. Sonmez and Switzer (2013) apply matching with contracts to match cadets with branches in US Military Academy. They present the cadet branch matching problem as a special case of matching with contracts (Hatfield and Milgrom, 2005) model and prove the application of the model beyond traditional two-sided matching problems.

Resource allocation methods in manufacturing industry focus on scheduling jobs to machines in a manufacturing facility with objectives such as maximizing efficiency, minimizing completion time or meeting due dates. Refer to Ouelhadj and Petrovic (2009) and Liu et al. (2019) for a survey on resource allocation and dynamic scheduling in manufacturing systems. This stream of literature does not involve the complexities associated with a network of independent suppliers. Thekinen and Panchal (2017) study the application of several economic matching mechanisms in different manufacturing scenarios such as centralized / decentralized settings and determine the optimal mechanism for each setting. In later work, Thekinen et al. (2018) determine the optimal frequency for matching orders with suppliers in a stochastic environment. They however, apply the deferred acceptance algorithm, to a manufacturing problem setting which liberally assumes that preferences of a supplier over orders are independent, do not match over contracts and do not maximize a utility objective. This is never the case in a manufacturing context.

## 3 Problem Description

A marketplace consisting of a network of independent suppliers on one side and clients requesting services on the other is considered. Suppliers register their machines on the marketplace platform in order to sell their underutilized capacity and clients in need of manufacturing services place orders on the marketplace. A client and his/her order is represented as an order in the rest of this paper. Orders, based on their attributes (such as required material, resolution and delivery date) are compatible with specific suppliers. The compatible set of orders and suppliers changes dynamically over time as new orders arrive and supplier capacities change. It is assumed that the platform has a planning period and it waits for the market to thicken and then performs the matching at the end of the planning period. The allocated orders and supplier capacities leave the marketplace and new orders and additional capacities arrive in each period. Remaining orders and supplier capacities either wait for the next matching cycle or perish. The mechanism of order allocation is a key design parameter of the marketplace platform.

Let $S$ be a set of suppliers and $s_j$, j $\in \{1, 2, \ldots, n\}$ represents the $j_{th}$ supplier. Let $D$ be a set of orders and $d_i$, i $\in \{1, 2, \ldots, m\}$ represents $i_{th}$ order. Orders arrive in the marketplace with a rate $\lambda$ per period and supplier reported capacities also arrive on the marketplace with a rate $\lambda_j$ hours per period for $j_{th}$ supplier. Suppliers communicate their capacities for the next $q$ periods to the platform. An order consists of the following attributes: 3D design model to be fabricated, required material, due date, preferred 3D printing process and resolution required. A supplier's service listing consists of the following attributes: type of machine (3D printing process), available materials, printable resolutions and available capacities. A supplier can accept multiple orders depending on its available capacity, the production time required for the orders and their due dates. For simplicity, we assume that each supplier has a single machine, although the solution can be easily extended to suppliers with multiple machines. The marketplace platform batches up orders to thicken the matching pool and allocates orders to suppliers at the end of every period.

Matching is performed by the platform based on the preferences of orders and suppliers over contracts. A contract $(i, j, c)$ consists of an order $i$, a supplier $j$ and terms $c$ such as material, 3D design, resolution, due date and price. An order supplier pair $(i, j)$ can have multiple contracts with different terms such as different prices, due dates or materials. Orders prefer certain attributes in contracts based on which it is determined the contract that has a higher preference over another. For instance, an order might prefer a contract with a supplier closer to its location which can deliver the order at a certain price and by a certain due date. Similarly, suppliers have preferences over terms of the contracts. For instance, a supplier might prefer a contract over another based on required material or customer type. Some of these attributes can be flexible, for example price, while some of them can be inflexible, for example 3D Design. Often there can be conflicting choices for an order or a supplier to make. For example a scenario where an order has to choose between a geographically closer supplier charging a higher price and a farther supplier charging a lower price. In scenarios where decisions have to be based on multiple criteria, expected utility theory, elaborated in Appendix A, can be used to rank alternatives. It has been used for multiple criteria decision making (Sanayei et al., 2008; Fernandez et al., 2005; Keeney and Raiffa, 1993; Min, 1994) in a variety of problems such as supplier selection. The expected utility theory is used to quantify and rank the contracts in the order of preference. Each quantified contract assigns a utility value to the order and supplier.

## 3.1 Stability in MaaS Marketplace

A matching $M$ in a MaaS marketplace is a many to one allocation of contracts between orders and suppliers where each order is assigned to a single supplier through a contract and each supplier can be assigned to multiple contracts each with a different order. A mechanism is a procedure of selecting a matching for each problem instance. The mechanism for matching in this paper considers the notion of stability to allocate contracts. Stability for a MaaS marketplace is defined in two contexts. First, it is defined by the size of coalition of participants blocking the match. In this context, the concepts of *pairwise stability* and *group stability* are considered. An order supplier pair $(i,j)$ with an acceptable contract $(i,j,c)$ is termed as a blocking pair when it meets the following criteria: order $i$ is either unassigned or strictly prefers $(i,j,c)$ over its assigned contract in $M$ and supplier $j$ is either underutilized (i.e. it can accept $(i,j,c)$ in addition to the contracts assigned in M) or it strictly prefers $(i,j,c)$ over one or more of its assigned contracts in $M$. A matching in which there is no blocking pair is pairwise stable. In case of non-strict preferences, the above definition of pairwise stability is termed as weak stability by Irving (1994). The term pairwise stability refers to weak pairwise stability for the rest of this paper. Group stability, a stronger concept of stability, ensures that a matching is not blocked by a coalition of participants of any size. In a MaaS marketplace, a group stable matching will ensure that there is no blocking coalition of size $n$ with 1 supplier and $(n-1)$ orders all of which would strictly prefer the coalition over their assigned matches. The coalition would consist of $(n-1)$ contracts each having the same supplier and a different order. Since a stable matching is not always guaranteed for a matching instance (with incomplete, non-strict and interdependent preferences) in a MaaS marketplace, this work considers approximately stable matchings which maximize the degree of stability i.e. minimize the number of blocking groups in a matching. There can also exist multiple approximately stable matchings with same number of blocking groups but with different total utilities. Therefore, it is reasonable to find an approximately stable matching with the highest total utility.

The second context which defines the concept of stability is the temporal dimension of matching. Since orders arrive on the platform continuously and supplier capacities change over time, the matching needs to be performed repetitively. In these dynamic settings, two notions of stability can be defined: *transient and posterior stability*. Transient stability is enforced in each matching cycle considering information (orders and supplier capacities) known in that cycle. However, new participants arriving in the next period can create blocking pairs/groups with participants in a transient stable match from the current period. When stability is enforced, considering information (orders and supplier capacities) received in two consecutive periods, it is called posterior stability. Posterior stability is achieved when participants in a matching cycle wait for an additional period for new participants to arrive and still fail to find a better match.

## 3.2 Impact of Stability

If the platform does not enforce stability i.e. elimination of blocking pairs/groups in the matching mechanism, it can find a matching which maximizes the total utility received by all participants. This matching is termed as maximum weight or socially optimal matching. Such a matching maximizes social welfare but might be unstable. Participants will have incentive to deviate from their assigned matches and create blocking pairs/groups. Enforcing stability would prevent creation of blocking pairs/groups but it can result in degradation of total utility achieved. This degradation can be measured by comparing the stable and socially optimal matchings for a matching instance. As defined by Anshelevich et al. (2013), the ratio of the highest and lowest total utility of the stable matchings to the total utility of the socially optimal matching is termed as price of stability and price of anarchy, respectively. Price of stability and price of anarchy quantifies the loss of social welfare because of enforcing stability in the system.

## 4 Proposed Matching Solutions

In this section, three formulations are proposed for establishing matching solutions in a MaaS marketplace. First, a maximum weight or socially optimal matching is formulated. This formulation maximizes the social welfare (collective value which all participants receive from the matching) and provides an upper bound on the total utility which can be achieved in a given matching instance. Second, a mathematical programming formulation which maximizes the social welfare while enforcing stability (individual participants getting their maximum possible value) is formulated and third, an approximately stable solution based on an extension of the Gale Shapley algorithm is proposed. Such an extension is known as a cumulative offer process (Hatfield and Milgrom, 2005).

### 4.1 Maximum Weight (MW)/ Socially Optimal Formulation

Consider a bipartite asymmetric edge-labeled multigraph $G$ with orders $i \in D$ and suppliers $j \in S$ as vertices. The contracts between feasible matches are represented as edges $(i, j, c)$ of $G$ where $c$ represents the terms of the contract between $i$ and $j$. Let $E$ be the set of all feasible contracts between all orders and suppliers in $G$. Let $X_{ijc}$ be a binary decision variable representing a contract $(i, j, c) \in E$ which is 1 if the contract is accepted and 0 otherwise. Let $C_{ij} \subseteq E$ represents the set of contracts between order $i$ and supplier $j$ and $u_{ijc}$ represents the total utility gained by order $i$ and supplier $j$ from contract $c \in C_{ij}$. Note that here with slight abuse of notation $(i, j, c)$ and $c$ have been used interchangeably. $p_{ij}$ represents the production time in hours required to manufacture the order $i$ (due in period $q$) on supplier $j$'s machine and is same for all contracts in $C_{ij}$. $h_{jq}$ represents supplier $j$'s available cumulative capacity in hours up to period $q$ and $q \in Q$ represents the set of periods in which the orders are due. Let $C_{ijq} \subseteq C_{ij}$ represents the set of contracts between order $i$ and supplier $j$ due on or before period $q$. The following matching formulation maximizes the total utility gained by all suppliers and orders in the platform.

$$\max \sum_{(i,j,c) \in E} u_{ijc} X_{ijc} \quad (1)$$

$$s.t. \sum_{j \in S} \sum_{c \in C_{ij}} X_{ijc} \leq 1 \quad \forall \ i \in D \quad (2)$$

$$\sum_{i \in D} \sum_{c \in C_{ijq}} p_{ij} X_{ijc} \leq h_{jq} \quad \forall (j \in S, q \in Q) \quad (3)$$

$$X_{ijc} \in \{0, 1\} \quad \forall \ (i, j, c) \in E \quad (4)$$

Constraints (2) are matching constraints which ensures that a single contract is accepted for each order. Constraints (3) are the feasibility constraints which ensures that each supplier has the capacity to manufacture the accepted orders before they are due in period $q$. For each supplier and each period, these constraints ensure that the time required to manufacture the orders due up to that period is less than or equal to the cumulative capacity available with the supplier up to that period.

## 4.2 Maximum Weight Approximately Stable (MWAS) Formulation

This formulation finds a matching with the largest utility and highest degree of stability i.e. the minimum number of blocking groups. The graph $G$ in the MW formulation is modified as follows. A supplier can accept contracts from multiple orders based on its available capacity and therefore has preferences over combination of contracts. Therefore, a new set of vertices $I$ which are a combination of individual order $i$ are added to $G$. The graph $G'$ with the expanded set of vertices $D'$ where $D \subseteq D'$ contains all possible combinations of orders for each supplier. The vertices $S$ on the supplier side remain same as in $G$. With single order vertices $D$ in $G$, the number of edges (individual contracts $c \in C_{ij}$) between a supplier order pair $(i, j)$ is equal to the number of contracts between them. The number of edges (combinations of contracts $C \in C_{Ij}$) between a combined order vertex $I$ with $n$ orders and a supplier $j$ is $\binom{m}{1}^n$ provided there are $m$ contracts between each order $i \in I$ and supplier $j$. With the expanded set of vertices, an order $i$ can now be associated with multiple vertices in $D'$. $D_i \subseteq D'$ contains all vertices associated with order $i$. $(I, j, C) \in E'$ represents the set of edges in $G'$. $(I, j, C)$ is a unique combination of individual contracts $(i, j, c)$ between all $i \in I$ and supplier $j$ and its weight $u_{IjC}$ is equal to the sum of utilities gained by all orders $i \in I$ and supplier $j$ from these contracts.

Since suppliers have limited capacity, the size of the graph can be reduced by removing the infeasible nodes in $D'$ and infeasible edges in $E'$. First, all combined order vertices $I \in D'$ which are not feasible for a supplier are identified. Given a set of individual order vertices which are feasible for a supplier, the size of the largest feasible combined vertex is found by the following. $max \ \sum_{i \in J} X_i$ subject to $\sum_{i \in J_q} p_{ij} X_i \leq h_{jq} \forall \ q \in Q$ where $i \in J$ denotes the set of individual order vertices for supplier $j$, $i \in J_q$ denotes the set of vertices in $J$ with due date upto period $q$ and $q \in Q$ represents the set of periods in which the orders $i \in J$ are due. $p_{ij}$ represents the production time in hours required to manufacture the order $i$ on supplier $j$'s machine and $h_{jq}$ represents supplier $j$'s available cumulative capacity in hours up to period $q$ starting from the current period. $X_i$ represents a decision variable which is 1 if vertex $i$ is accepted and 0 if it is rejected. This formulation provides an upper bound on the number of orders a supplier can serve given a set of orders and its available capacities. It is solved for each supplier $j \in S$ and any nodes in $D'$ with size greater than the upper bound are removed from $G'$ for each supplier. Second, for the remaining nodes and edges in $G'$, feasible edges in $E'$ are identified using constraints (3) and infeasible edges are removed from $G'$. The MWAS formulation is executed in two steps. In the first step, a matching which minimizes the number of blocking groups is found as follows:

$$min \sum_{I,j,C \in E'} Y_{IjC} \tag{5}$$

$$s.t. \sum_{j \in S} \sum_{C \in C_{Ij}} \sum_{I \in D_i} X_{IjC} \leq 1 \quad \forall \ i \in D \tag{6}$$

$$\sum_{I \in D'} \sum_{C \in C_{Ij}} X_{IjC} \leq 1 \quad \forall \ j \in S \tag{7}$$

$$Y_{IjC} + \sum_{i \in I} \sum_{I'j'C' \succcurlyeq_i IjC} X_{I'j'C'} + \sum_{I'jC' \succcurlyeq_j IjC} X_{I'jC'} + X_{IjC} \geq 1 \ \forall \ (I, j, C) \tag{8}$$

$$X_{IjC} \in \{0, 1\}, \quad Y_{IjC} \in \{0, 1\} \quad \forall \ (I, j, C) \tag{9}$$

Binary decision variables $X_{IjC}$ and $Y_{IjC}$ represent a combination of contracts $(I, j, C)$ between designers $i \in I$ and supplier $j$. $X_{IjC}$ is 1 if $(I, j, C)$ is accepted and 0 otherwise. $Y_{IjC}$ is 1 if $(I, j)$ forms a blocking group with combination of contracts $C$ and 0 otherwise. Constraints (6) ensure that each order is accepted only once and Constraints (7) ensure that each supplier accepts only one possible combination of contracts. Constraints (8) ensure for each $(I, j, C) \in E'$ that either edge $(I, j, C)$ is accepted or supplier $j$ is assigned another set of contracts $(I'jC')$ which it prefers at least as well (denoted by $\succcurlyeq$) as $(I, j, C)$ or at least one of the orders $i \in I$ is assigned $(I', j, C')$ which it prefers at least as well as $(I, j, C)$. If none of the above conditions are met, then $(I, j)$ is considered a blocking group with combination of contracts $C$ and $Y_{IjC}$ takes the value 1. In summary, either $(I, j, C)$ is accepted or at least one of the orders or suppliers is assigned a match which it prefers as well or $(I, j)$ becomes a blocking group with combination of contracts $C$. These constraints will ensure that if there is a group each of whose participant is either unmatched or strictly prefers being in the group over its assigned match, it will be considered a blocking group. The objective function in (5) minimizes the number of blocking groups and provides a lower bound (LB) on the number of blocking groups in a matching instance.

In the second step, the objective function is changed to $max \sum_{I,j,C \in E'} u_{IjC} X_{IjC}$ and a new constraint $\sum_{I,j,C \in E'} Y_{IjC} \leq \text{LB}$ is added in addition to constraints 6, 7 and 8. The formulation in the second step finds a matching with the highest utility among the matchings with the number of blocking groups restricted to lower bound found in the first step. The objective function in the second step can be modified to $min \sum_{I,j,C \in E'} u_{IjC} X_{IjC}$ to find a matching with minimum utility and to $max \sum_{I,j,C \in E'} X_{IjC}$ to find a matching with maximum cardinality. The minimum weight approximately stable matching provides a lower bound for the utility achieved and the maximum cardinality approximately stable matching maximizes the number of matches in any approximately stable matching.

### 4.3  Approximately Stable (AS) Formulation

This section proposes an extension of the Gale-Shapley algorithm. The Graph $G$ in the maximum weight formulation (Section 4.1) is considered where we have vertices as the set of orders $i \in D$, suppliers $j \in S$ and edges $(i, j, c) \in E$. The algorithm executes as described in Figure 1. Each order $i \in D$ proposes to the supplier $j \in S$ in its highest ranked contract i.e. the contract with the largest order utility in its list of contracts. Since all edges $E$ in $G$ are feasible, the contracts are tentatively accepted by the suppliers. Note that even though the individual edges in $G$ are feasible, a supplier needs to evaluate which combination of edges/ contracts it can fulfill considering its capacity and the due date of contracts. This results in each supplier having a list of contracts accepted tentatively. The choice function (elaborated in the next paragraph), is then used to evaluate the tentatively accepted contracts for each supplier and it then selects the subset it can fulfill while achieving highest utility. The remaining contracts are rejected. This completes the first round of tentative acceptance within a given period. The second round processes the orders in rejected contracts from the previous round. The rejected orders now propose to their second ranked contract and suppliers now choose from the combination of tentatively accepted orders from previous rounds and the current round. This process repeats until there are no rejected orders or all the contracts of rejected orders have been explored ensuring that the algorithm will always terminate for any matching instance.

---

**Approximately Stable Matching Algorithm**

**set** all orders as rejected orders

**set** an empty list for tentatively assigned contracts for each supplier

**input:** list of contracts for each order ranked by its utility

**while** any order *d* is rejected **do**:

    **if** set of contracts for each rejected order is empty: **break**

    **tentatively assign** rejected orders to suppliers in their first ranked contract

    **add** assigned contracts to suppliers' list of tentatively assigned contracts

    **delete** first ranked contract from each order's list of contracts

    **for** each supplier *s* **do**:

        **apply** choice function to accept a subset of contracts from tentatively assigned contracts

        **del** rejected contracts from its list of tentatively assigned contracts

        **append** orders from rejected contracts to list of rejected orders

    **end for**

**end while**

---

**Figure 1:** Algorithm for Approximately Stable Formulation

The choice function chooses the set of contracts which can be processed by their due date and maximize the utility of a supplier $j \in S$. The maximum weight formulation reduced to a single supplier is used as a choice function. The objective function reduces to $\max \sum_{(i,c) \in J} u_{j_ic} X_{ic}$ where $u_{j_ic}$ is the utility gained by supplier $j$ from a contract $(i, j, c)$, constraints (2) are no longer required and constraint (3) reduces to $\sum_{(i,c) \in B} p_{ij} X_{ic} \leq h_{jq} \forall q \in Q$ where $(i, c) \in J$ denotes the set of all contracts tentatively assigned to supplier $j$ with no two contracts having the same order $i$ and $(i, c) \in B$ denotes the set of contracts in $J$ with due date upto $q$. $h_{jq}$ represents supplier $j$'s available cumulative capacity in hours up to period $q$ and $q \in Q$ represents the set of periods in which the tentatively assigned contracts are due.

Though MWAS formulation will always find a stable matching if one exists, this algorithm (AS) may fail to find a stable matching the reasoning for which is explained as follows. If the choice function for suppliers satisfies a substitute condition, this algorithm becomes pairwise and group stable as demonstrated by Hatfield and Milgrom (2005). The substitute condition states that any contract $(i, j, c)$ that is rejected by a supplier from a smaller set $X$ is also rejected from any larger set $X'$ that contains $(i, j, c)$. The choice function described above fails the substitute condition because of interdependent preferences of suppliers over contracts in the MaaS marketplace problem setting. For instance, consider the following four contracts each with a different designer and same supplier with production time and supplier utility tuples as follows: $c_1$: (8.1, 0.95), $c_2$: (4.6, 0.80), $c_3$: (4.1, 0.72) and $c_2$: (4.4, 0.78). All the contracts are due in same period and supplier has a cumulative capacity of 9 hours until the due date. From a smaller set of $(c_1, c_2)$, $c_2$ would be rejected. However, from any larger set of 3 or 4 contracts, $c_2$ will always be chosen. The impact of choice function failing the substitute condition would depend on the sequence of arrival of contracts in the algorithm. If $c_1$ comes first, it would result in violation of stability however, if it comes last, it would not impact the stability of matching. The stability/ existence of blocking pairs/groups in a matching instance can be arbitrary in this case. Therefore, in Section 7, the impact of the failure of this substitute condition on matching stability is evaluated empirically in a MaaS marketplace. Considering the marginal impact on stability found in results, this algorithm is termed as "Approximately Stable" for a MaaS marketplace problem setting. If suppliers specify predetermined number of acceptable contracts and preferences over contracts are independent, as assumed in Thekinen and Panchal, 2017, the AS algorithm described above becomes pairwise and group stable. However, the number of contracts acceptable to a supplier depends on the attributes of contracts and cannot be predetermined.

## 5  Stability of the Proposed Matching Solutions

### 5.1  Transient Stability

Recall that a blocking pair (BP) is formed by an order-supplier pair $(i,j)$ with terms of contract $c$ each of whom is either unassigned (underutilized in case of suppliers) or strictly prefers $(i,j,c)$ over its assigned contract(s). Let $u_{i_jc}$ is the utility received by order $i$ and $u_{j_ic}$ is the utility received by supplier $j$ from $(i,j,c)$ where $u_{ijc} = u_{i_jc} + u_{j_ic}$. For $(i,j)$ to qualify as a blocking pair with terms of contract $c$, $u_{ijc} > u_{i_{j'}c'}$ or $i$ should be unassigned and $\sum_{J'} u_{j_{i'}c'} > \sum_{J} u_{j_{i'}c'}$ where $i$ has been matched to $(i,j',c')$, $J$ is the set of contracts to which $j$ has been matched and $J'$ which includes $(i,j,c)$ is the set of contracts chosen by the supplier from the set $J \cup (i,j,c)$. The set of contracts $J'$ are identified using the choice function described in Section 4.3. If $J \subset J'$ then the supplier $j$ is considered underutilized in the blocking pair $(i, j)$. If $i$ is unassigned, it is considered an unmatched participant in the blocking pair $(i, j)$. If the order in the blocking pair is unmatched and the supplier is underutilized, it is termed as an available blocking pair. The MW matching and MWAS matching will always match an available blocking pair since it would always strictly increase the total utility. The AS matching can have an available blocking pair depending on the sequence of arrivals of contracts as described in Section 4.3. Similar to the procedure described for pairwise stability, blocking groups (BG) can be identified. For a group $(I,j)$ to be a blocking group, each $i \in I$ should have a feasible contract with $j$ and terms $c$ where $u_{ijc} \geq u_{i_{j'}c'}$ or $i$ should be unassigned and $\sum_{J'} u_{ijc} > \sum_{J} u_{j_{i'}c'}$ where $i \in I$ has been matched to some $(i,j',c')$, $J$ is the set of contracts to which $j$ has been matched and $J'$ is the blocking group $(I,j)$. Notice that this definition allows $J'$ to have some of the contracts assigned to $j$ in $J$. These contracts which are in $J \cap J'$ are indifferent between their current assignment and the blocking group. The contracts in the set $J' \setminus J$ strictly prefer the blocking group over their assigned match. If $J \subset J'$ then the supplier $j$ is considered underutilized in the blocking group $(I,j)$. If an $i$ in $I$ is unassigned, it is considered an unmatched participant in the blocking group $(I, j)$. If all orders in the blocking group are unmatched and the supplier is underutilized, it is termed as an available blocking group. The MW matching and MWAS matching formulation will always match an available blocking group since it would always strictly increase the total utility. The AS matching formulation can have an available blocking group depending on the sequence of arrivals of contracts as described in Section 4.3.

### 5.2  Posterior Stability

Recall that posterior stability considers the orders and supplier capacities received during two consecutive matching periods. The procedure to identify the blocking pairs and groups remains same as in the case of transient stability with two key modifications. First, the set of participants which are considered to identify the blocking pairs is wider with participants from two consecutive periods being considered. Second, the choice function described in Section 4.3 is modified to maximize $\sum_{(i,c) \in J} u_{j_ic} X_{ic}$ subject to $\sum_{(i,c) \in B_{tq}} p_{ij} X_{ic} \leq h_{tq} \forall q \in Q_t$ , $\sum_{(i,c) \in B_{(t+1)q}} p_{ij} X_{ic} \leq h_{(t+1)q} \forall q \in Q_{t+1}$ and $\sum_{(i,c) \in B_q} p_{ij} X_{ic} \leq h_{tq} \forall q \in Q_t \cup Q_{t+1}$ where $t$ and $t+1$ denotes the two consecutive periods in which orders arrive and $Q_t$ and $Q_{t+1}$ represents the set of periods in which the arriving orders in these periods are due respectively. $(i,c) \in$

$B_{tq}$ represent all orders which arrive in period $t$ and are due in periods upto $q$, $(i, c) \in B_{(t+1)q}$ represent all orders which arrive in period $t+1$ and are due in periods upto $q$. $(i, c) \in B_q$ represents the set of orders which are due in periods upto $q$. $h_{tq}$ and $h_{(t+1)q}$ represents the cumulative capacity of supplier $j$ from period $t$ and $t+1$ to $q$ respectively. $(i, c) \in J$ represents all orders arrived in both periods $t$ and $t$ +1.

## 6 Empirical Simulation Environment

The simulation environment consists of a marketplace with a network of 100 suppliers with each assumed to have one machine. The machines consist of 3D printers based on Fused Deposition Modelling (FDM), Selective Laser Sintering (SLS), Stereolithography (SLA) and Material Jetting. The machine specifications include the materials and resolution at which it can process those materials. Materials include Polylactic Acid (PLA), Nylon, Polycarbonate (PC), Acrylonitrile styrene acrylate (ASA), Resins, Thermoplastic polyurethane (TPU), Aluminum and Steel alloy. Machines, resolution range and materials were taken from supplier profiles from the 3D Hubs marketplace dataset used in Pahwa and Starly (2019) or from machine specifications available on vendor site. Suppliers list their machine capacities of up to 6 hours per period on the platform for 4 periods into the future. Length of one period is considered to be 12 hours. The middleware platform has a reputation system where orders rate suppliers based on their service quality. Supplier ratings vary on a scale of 1 to 5 with 5 being the best rating. Though 3D Hubs dataset had majority of supplier listings (84%) as FDM machines, we consider FDM having 50% of the listings, SLA 15%, material jetting 15%, polymer based SLS 15% and metal based SLS 5%. This provides a reasonable mix of all types of machines in a MaaS marketplace.

Orders that arrive on the marketplace consists of attributes such as 3D model information, material required, desired 3D printing process, required resolution and a due date. Due dates for the orders vary from 2 periods up to 7 periods from the period in which the order arrived. Different probability distributions are used to generate these attributes. For instance, 3D printing processes for the arriving orders have a probability mass function where the probability of FDM order arrival is 0.5, SLA 0.15, material jetting 0.15 and SLS 0.2 each. Average production time (production and post processing) required for an order arriving on the platform is 5.35 hours and orders have a Poisson arrival with a rate ($\lambda$) of 100 orders per period. An order is considered feasible for a supplier based on four machine attributes – 3D printing process, material, resolution and available capacity. Contracts are generated for each order and its feasible suppliers as represented by $(i, j, c) \in E$ (section 4.1). Additionally, feasible combinations of contracts are generated for suppliers as represented by $(i, j, c) \in E'$ (section 4.2). A contract consists of order attributes (material, process, resolution and due date), a supplier, and a price. Price is considered to be the flexible term in the contract and up to two contracts are generated between a supplier-order pair.

Suppliers' value three attributes: the revenue to be obtained from the order, required material and urgency. The higher the revenue, the higher is the utility for suppliers. The farther the due date, the more available time available to fulfill order leading to higher utility. The utility from the material depends on the supplier's inventory. If the material is readily available with the supplier, the utility is higher. Orders value a supplier's size, rating, location and price to pay. Suppliers are categorized as small, medium or large scale and orders specify their preferences over size. Some orders prefer to work with large firms whereas some prefer small scale workshops. Orders prefer monotonically decreasing price for the order. There are five levels of supplier ratings with a higher rating meaning a better supplier and orders have a monotonically increasing preference for supplier rating. Location data for suppliers is used from 3D Hubs dataset used in Pahwa and Starly (2019). Suppliers and orders are assigned a location (GPS coordinates) in the US and the distance between them is calculated. The closer a supplier is to an order, the higher its utility.

Incoming orders are accumulated in each period and matching is performed at the end of the period. Matched orders leave the marketplace and rejected orders remain in the platform to be considered in the next period until their due date. Supplier capacities are updated considering the assigned orders. New supplier capacities and new orders arrive in the next period. New orders along with the remaining orders from the previous periods are matched at the end of this period. This process repeats and the simulation is conducted over a length of 15 periods.

## 7 Results and Analysis

This section presents the results and analysis from simulation experiments conducted to evaluate the impact of stability in a MaaS marketplace. The matching solutions presented in Section 4 were implemented in using Python 3.7 and Gurobi 8.1. The experiments were performed on 3.70GHz Xeon CPU with 24 GB RAM. Following sections present the results from the simulation experiments.

### 7.1 Influence of Enforcing Stability on System Performance

The following metrics are used to quantify the impact of enforcing stability on platform's performance:

**Impact of stability:** This is the ratio of total utility achieved by stable matching to the total utility achieved by maximum weight matching. This represents the price of stability and anarchy for maximum and minimum weight approximately stable matching, respectively.

**Average participant utility:** This is the average utility gained by each matched participant evaluated separately for orders and suppliers. For suppliers, this represents average utility attained per matched order.

**Matched participants:** This is the proportion of matched participants evaluated separately for suppliers and orders.

**Average participant rank:** This is the average percentile rank of a matched contract evaluated separately for suppliers and orders.

Table 1: System performance metrics for proposed solutions with $\lambda = 100$; Averages over 5 instances of 15 periods each (95% confidence intervals (CI))

|  | AS | MWAS | MW |
|---|---|---|---|
| Impact of stability | 0.947 (0.007) | 0.954 (0.004) | 1 |
| Average order utility | 0.679 (0.003) | 0.681 (0.003) | 0.73 (0.003) |
| Average supplier utility | 0.681 (0.006) | 0.679 (0.006) | 0.676 (0.006) |
| Matched orders | 0.731 (0.016) | 0.738 (0.015) | 0.748 (0.015) |
| Matched suppliers | 1 | 1 | 1 |
| Average order rank | 0.282 (0.012) | 0.279 (0.009) | 0.234 (0.006) |
| Average supplier rank | 0.413 (0.026) | 0.422 (0.026) | 0.465 (0.016) |

Table 1 presents the results for the metrics defined above. In addition to the three proposed solutions, the minimum weight and maximum cardinality approximately stable matchings are also evaluated. The price of anarchy and price of stability are 0.917 and 0.954 respectively. This demonstrates that any approximately stable matching instance where the objective is not to maximize the utility will not be worse than 0.917 of the socially optimal. The AS matching with 0.947 impact of stability is very close to the MWAS matching. The results are in concurrence with Anshelevich et al. (2013) where they show that in case of asymmetric edge-labeled graphs, theoretically, the impact of stability can be arbitrarily bad. However, in simulated matching instances under different utility distributions, they show that the price of anarchy is above 0.85. Owing to a capacity constrained environment, all suppliers in the system are matched with at least one order in all three formulations. All three formulations and the maximum cardinality approximately stable matching which provides an upper bound on the number of matched participants also matches almost same number of participants (74% orders and 100% suppliers) establishing that in a capacity constrained environment all stable matches will assign almost same number of matches. Orders receive a higher utility in the MW matching leading to a lower average percentile rank compared to the other two solutions.

The computation time per matching instance (average over 15 periods) for AS matching is 9.4 seconds compared to 124.5 seconds for MWAS matching. The large number of stability constraints (equal to number of combinations of feasible contracts between suppliers and orders) in the MWAS formulation make it computationally expensive. For larger instances, the graph $G'$ can be divided into subgraphs based on parameters such as machine types or location for computational efficiency. The subgraphs can still be large depending on the size of the network of suppliers in the platform. Algorithms such as Column Generation (Dantzig and Wolfe, 1960) can be developed to efficiently solve this formulation for larger instances. Since the AS matching performs equally well in terms of system performance and provides a significantly faster solution, it is recommended for solving larger instances of the problem.

### 7.2 Transient Stability of Proposed Solutions

The following metrics are used to evaluate transient stability of the solutions proposed in Section 4:

**Participants in blocking pairs/groups:** This is the total number of unique participants in blocking pairs/groups divided by total number of participants evaluated separately for orders and suppliers.

**Average blocking pairs/groups per participant:** This is the total number of blocking pairs/groups divided by number of unique participants in blocking pairs/groups evaluated separately for orders and suppliers. For suppliers, it is further averaged over the number of periods.

**Unmatched participants in blocking pairs/groups:** This is the number of unique unmatched (underutilized in case of suppliers) participants in blocking pairs/groups divided by the total number of unique participants in blocking pairs/groups evaluated separately for orders and suppliers.

**Available blocking pairs/groups:** This is the number of available blocking pairs/groups divided by the total number of blocking pairs/groups in a matching.

**Average participant utility gain:** This is the difference between the utility received from the blocking pair/group match and utility received from assigned match divided by the utility received from assigned match averaged over all participants and evaluated separately for orders and suppliers.

**Average size of a blocking group:** This represents the average number of participants in a blocking group. The number of participants in a blocking pair is always two. However, blocking groups can have a varied number of orders associated with a single supplier.

Table 2: Transient stability metrics for proposed solutions with λ = 100; Averages over 5 instances of 15 periods each (95% CI)

|  | AS | | MWAS | | MW | |
| --- | --- | --- | --- | --- | --- | --- |
|  | BP | BG | BP | BG | BP | BG |
| Orders in BP/ BG | 0.006 (0.003) | 0.069 (0.011) | 0.001 (0.001) | 0.001 (0.002) | 0.44 (0.015) | 0.471 (0.014) |
| Suppliers in BP/ BG | 0.074 (0.054) | 0.312 (0.083) | 0.01 (0.018) | 0.01 (0.018) | 0.982 (0.02) | 0.984 (0.017) |
| Avg. BP/ BG per order | 1.025 (0.069) | 2.577 (0.435) | 0.4 (0.68) | 0.4 (0.68) | 2.84 (0.128) | 3.437 (0.306) |
| Avg. BP/ BG per supplier | 0.09 (0.018) | 0.289 (0.075) | 0.027 (0.045) | 0.027 (0.045) | 1.272 (0.08) | 1.424 (0.133) |
| Unmatched orders | 0.129 (0.174) | 0.326 (0.114) | 0 | 0 | 0.251 (0.033) | 0.241 (0.033) |
| Underutilized suppliers | 0.143 (0.113) | 0.03 (0.023) | 0.067 (0.185) | 0.067 (0.185) | 0.103 (0.065) | 0.103 (0.065) |
| Available BP/ BG | 0.079 (0.18) | 0 | 0 | 0 | 0 | 0 |
| Avg. order gain | 0.234 (0.19) | 0.454 (0.145) | 0.045 (0.08) | 0.032 (0.059) | 0.33 (0.025) | 0.3 (0.02) |
| Avg. supplier gain | 0.163 (0.132) | 0.416 (0.133) | 0.078 (0.205) | 0.06 (0.153) | 0.144 (0.013) | 0.147 (0.017) |
| Avg. size of BP/BG | 2 | 3.01 (0.023) | 2 | 0.967 (1.645) | 2 | 2.154 (0.036) |

Table 2 presents the results for transient stability metrics. The AS matching has only a few participants in blocking pairs and the number blocking pairs in which these participants are present are low. The performance is relatively worse in terms of group stability which is expected as it is a stronger notion of stability. The MWAS matching performs the best in terms of stability which is expected since this formulation minimizes the number of blocking groups. The MW matching is highly unstable as expected and performs the worst with almost half of orders and all suppliers in the blocking pairs and groups. These participants also have relatively much higher number of opportunities to form matches outside the system compared to the other two solutions. In Appendix B, the impact of order arrival rate is studied on the performance of the system in terms of utility and stability.

### 7.3 Posterior Stability of Proposed Solutions

Table 3 presents the results when transient stable matches are evaluated for posterior stability. Participants in a specific period will have additional match opportunities owing to waiting participants from the previous period and if they choose to wait for the new participants arriving in the next period. Therefore, this represents the worst case scenario in a hypothetical situation where all participants will have up to thrice the match opportunities they receive in a transient stable matching. Considering this, period wise matching solutions are expected to perform significantly worse in terms of posterior stability as compared to transient stability. The results in Table 3 demonstrate that in terms of posterior stability as well, both AS and MWAS matching perform significantly better as compared to the MW matching. Though all three matchings have almost equal number of participants in blocking pairs/groups, the number of opportunities to form blocking pairs/groups available to participants in AS and MWAS matchings are significantly lower compared to the MW matching. The posterior stability results demonstrate the importance of dynamic matching where agents can choose to wait expecting better matches in upcoming periods as discussed in Akbarpour et al. (2017).

Table 3: Posterior stability metrics for proposed solutions with λ = 100; Averages over 5 instances of 15 periods each (95% CI)

|  | AS | | MWAS | | MW | |
| --- | --- | --- | --- | --- | --- | --- |
|  | **BP** | **BG** | **BP** | **BG** | **BP** | **BG** |
| Orders in BP/ BG | 0.705 (0.031) | 0.891 (0.016) | 0.705 (0.022) | 0.888 (0.011) | 0.756 (0.012) | 0.938 (0.004) |
| Suppliers in BP/ BG | 1 | 1 | 1 | 1 | 0.998 (0.006) | 0.998 (0.006) |
| Avg. BP/ BG per order | 4.007 (0.145) | 51.574 (15.85) | 4.039 (0.208) | 44.215 (11.896) | 7.16 (0.218) | 105.074 (24.474) |
| Avg. BP/ BG per supplier | 2.826 (0.287) | 19.71 (6.563) | 2.848 (0.305) | 17.565 (4.972) | 5.423 (0.333) | 43.115 (10.302) |

### 7.4 Influence of Switching Costs

The suppliers and orders in a blocking pair/group might not deviate from the match assigned by the platform if the benefit they are achieving by deviating is marginal. Also, to incentivize participants to accept its assigned match, the platform can impose a switching cost if a participant deviates from its assigned match. Therefore, it is important to analyze the benefit participants receive by forming blocking pairs/groups and deviating from the assignments in the proposed solutions.

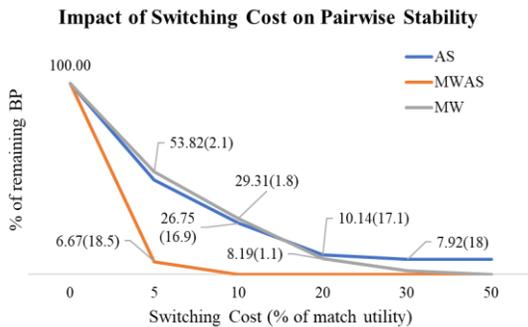

Figure 2: Impact of switching cost on pairwise stability; Averages over 5 instances of 15 periods each (95% CI)

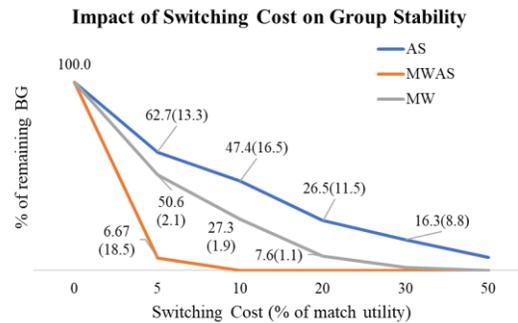

Figure 3: Impact of switching cost on group stability; Averages over 5 instances of 15 periods each (95% CI)

A multiplicative switching cost (Anshelevich et al., 2013) which is a proportion of the utility received by the participant in an assigned match is considered. Figure 2 and Figure 3 presents the impact of imposing switching costs on the number of blocking pairs and blocking groups for the proposed solutions. In the MW matching, the number of blocking pairs and blocking groups gradually decline as the switching cost increases reaching to almost zero at 30% switching cost. This demonstrates that the platform can enforce stability even with a MW matching by imposing a switching cost of at least 30% of achieved utility. The rate of decline of the blocking pairs and groups is almost same because the size of an average blocking group is 2.15 (from Table 2). The AS and MWAS matchings have significantly lesser number of blocking pairs/groups which varies highly among matching instances leading to wide confidence intervals for remaining blocking pairs/groups when the switching cost is increased gradually.

### 7.5 Impact of Unstable Matches over Time under Socially Optimal (MW) Allocation

In this section impact of unstable matches under the MW allocation is evaluated. When the matching is unstable, the participants can form matches outside the platform rejecting its allocation. Orders approach their known suppliers in the decreasing order of preference and if accepted, they reject the platform's allocation. To introduce this form of anarchy into the system, the first round of the AS algorithm is executed on the results of MW allocation with a key difference that these allocations of defying participants are not tentative but final. First a "Complete Access" scenario is considered where each order has access to the identities of all suppliers and second, a "Restricted Access" scenario is considered where orders only know the identities of suppliers they have transacted with in the past.

Figure 4 shows the impact of participants rejecting unstable matches on the total utility of the system. AS matching achieves 95.2% utility of the optimal allocation on average over a length of 100 periods whereas unstable matching under Complete Access achieves 74.16% of optimal utility. Under Restricted Access, the total utility remains close to the optimal allocation initially as fewer participants are able to reject the unstable matches. However, as the orders transact with more suppliers, their network of known suppliers grows and the total system utility almost reaches the level of complete access by the end of 100 periods. Figure 5 presents the percentage of defying participants under both scenarios. As the network of order grows under the restricted access scenario, the number of defying participants grows and almost reaches the level of defying participants under complete access. To prevent defying participants i.e. matches forming outside the platform, most, if not all MaaS platforms, hide the identity of the participants. These results demonstrate that the platform is better off over the long run being open and transparent. When the matching is stable, participants bypassing the platform is not a concern and the platform does not need to hide the identities and prevent access to the participants.

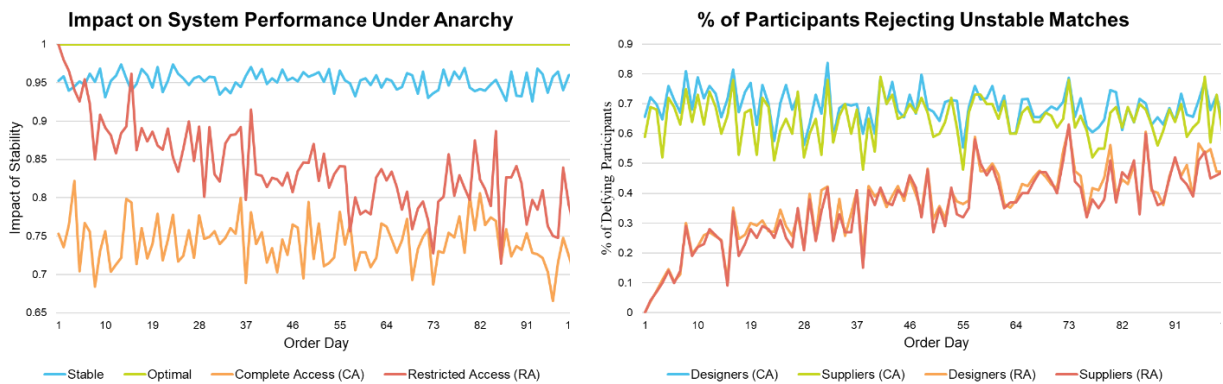

Figure 4: Impact on total utility when participants reject unstable matches

Figure 5: Percentage of participants rejecting unstable matches under restricted and complete access

## 8    Discussion and Future Opportunities

Online manufacturing services marketplaces have popped up in several major markets globally with service abilities in short-run part production. However, as the market grows with additional service offerings, demand naturally increases leading to the need for enhanced transparency. Most current MaaS platform restrict the identities of the participants. Our analysis and simulation shows that in the long term, restricting the identity of the participants can adversely affect the platform health due to the lack of transparency and utility obtained by its participants when actively being within the platform. Sustaining operations of the platform can only be ensured when its participants receive fair share value for the interaction offered through the platform. Stable bipartite mechanisms promote fairness and transparency through decentralization as the decision making is driven by the preferences of the participants. MaaS platforms can still be profitable when these marketplaces collect and store data on the suppliers and orders as related to collecting individual participants' preferences over time. Data-driven algorithms can be designed so that better contracts are automatically derived providing utility to both clients and suppliers. If such interaction value is provided, the participants do not necessarily have to by-pass the platform for conducting business.

On the technical side, computationally efficient approaches to solve matchings, particularly MWAS matching need to be developed. Additionally, matching frequency can impact the performance of the system as demonstrated by the results on posterior stability of matching algorithms. Participants can benefit from strategic decision making to enter into the matching pool or waiting for a better match (Akbarpour et al., 2017) in each matching period. Such strategic decision making can lead to posterior stability in matching, which is equally important for long-term stability of the marketplace. The utilities of participants over feasible matches change with time and waiting for a match can also be expensive for participants. These additional dimensions provide avenues for further research in this area.

## 9    Conclusion

In decentralized MaaS marketplaces where manufacturing service requests from clients are assigned to suppliers by the marketplace platform, the participants (client and suppliers) can reject the assignment of the platform and find a better assignment on their own. Therefore, it is important for the platform to consider the notion of stability while matching the participants. This work introduces stable matching methods for a MaaS marketplace and empirically evaluates them for different criteria of stability. The results demonstrate that enforcing stability introduces a slight loss in the total utility achieved by the platform. However, it ensures a sustainable marketplace in the long run. Among the proposed stable solutions, the AS matching is recommended. It performs equally well in terms of utility, slightly worse in terms of stability but is significantly more time efficient compared to the MWAS matching. In a networked environment, MaaS marketplaces have an increasingly important role to play, particularly in prototyping service scenarios and its potential role in enabling the production of heavily customized limited quantity products.

**References**
1. Abdulkadiroğlu, A., & Sönmez, T. (2003). School choice: A mechanism design approach. American economic review, 93(3), 729-747.
2. Abeledo, H., & Blum, Y. (1996). Stable matchings and linear programming. *Linear algebra and its applications*, *245*, 321-333.

3. Akbarpour, M., Li, S., & Oveis Gharan, S. (2017). Thickness and information in dynamic matching markets. Available at SSRN 2394319.

4. Anshelevich, E., Das, S., & Naamad, Y. (2013). Anarchy, stability, and utopia: creating better matchings. Autonomous Agents and Multi-Agent Systems, 26(1), 120-140.

5. Balinski, M., & Sönmez, T. (1999). A tale of two mechanisms: student placement. Journal of Economic theory, 84(1), 73-94.

6. Dantzig, G. B., & Wolfe, P. (1960). Decomposition principle for linear programs. Operations research, 8(1), 101-111.

7. Fernandez, M. G., Seepersad, C. C., Rosen, D. W., Allen, J. K., & Mistree, F. (2005). Decision support in concurrent engineering–the utility-based selection decision support problem. Concurrent Engineering, 13(1), 13-27.

8. Gale, D., & Shapley, L. S. (1962). College admissions and the stability of marriage. The American Mathematical Monthly, 69(1), 9-15.

9. Gale, D., & Sotomayor, M. (1985). Some remarks on the stable matching problem. *Discrete Applied Mathematics*, *11*(3), 223-232.

10. Hatfield, J. W., & Kojima, F. (2009). Group incentive compatibility for matching with contracts. Games and Economic Behavior, 67(2), 745-749.

11. Hatfield, J. W., & Kojima, F. (2010). Substitutes and stability for matching with contracts. Journal of Economic theory, 145(5), 1704-1723.

12. Hatfield, J. W., & Milgrom, P. R. (2005). Matching with contracts. American Economic Review, 95(4), 913-935.

13. Irving, R. W. (1994). Stable marriage and indifference. Discrete Applied Mathematics, 48(3), 261-272.

14. Iwama, K., & Miyazaki, S. (2008, January). A survey of the stable marriage problem and its variants. In *International conference on informatics education and research for knowledge-circulating society (ICKS 2008)* (pp. 131-136). IEEE.

15. Iwama, K., Miyazaki, S., Morita, Y., & Manlove, D. (1999, July). Stable marriage with incomplete lists and ties. In *International Colloquium on Automata, Languages, and Programming* (pp. 443-452). Springer, Berlin, Heidelberg.

16. Keeney, R. L., & Raiffa, H. (1993). Decisions with multiple objectives: preferences and value trade-offs. Cambridge university press.

17. Liu, T., Wan, Z., & Yang, C. (2019). The efficiency of a dynamic decentralized two-sided matching market. Available at SSRN 3339394.

18. Liu, Y., Wang, L., Wang, X. V., Xu, X., & Zhang, L. (2019). Scheduling in cloud manufacturing: state-of-the-art and research challenges. International Journal of Production Research, 57(15-16), 4854-4879.

19. Manlove, D. F., McBride, I., & Trimble, J. (2017). "Almost-stable" matchings in the Hospitals/Residents problem with Couples. Constraints, 22(1), 50-72.

20. Min, H. (1994). International supplier selection: a multi-attribute utility approach. International Journal of Physical Distribution & Logistics Management, 24(5), 24-33.

21. Ouelhadj, D., & Petrovic, S. (2009). A survey of dynamic scheduling in manufacturing systems. Journal of scheduling, 12(4), 417.

22. Pahwa, D., & Starly, B. (2019). Network-based pricing for 3D printing services in two-sided manufacturing-as-a-service marketplace. Rapid Prototyping Journal.

23. Roth, A. E. (1984). The evolution of the labor market for medical interns and residents: a case study in game theory. *Journal of political Economy*, *92*(6), 991-1016.

24. Roth, A. E. (2008). What have we learned from market design?. Innovations: Technology, Governance, Globalization, 3(1), 119-147.

25. Roth, A. E., Rothblum, U. G., & Vande Vate, J. H. (1993). Stable matchings, optimal assignments, and linear programming. *Mathematics of operations research*, *18*(4), 803-828.

26. Roth, A. E., & Sotomayor, M. (1992). Two-sided matching. Handbook of game theory with economic applications, 1, 485-541.

27. Rothblum, U. G. (1992). Characterization of stable matchings as extreme points of a polytope. *Mathematical Programming*, *54*(1-3), 57-67.

28. Sanayei, A., Mousavi, S. F., Abdi, M. R., & Mohaghar, A. (2008). An integrated group decision-making process for supplier selection and order allocation using multi-attribute utility theory and linear programming. Journal of the Franklin institute, 345(7), 731-747.

29. Sönmez, T., & Switzer, T. B. (2013). Matching with (branch-of-choice) contracts at the United States military academy. Econometrica, 81(2), 451-488.

30. Thekinen, J. D., Han, Y., & Panchal, J. H. (2018, August). Designing Market Thickness and Optimal Frequency of Multi-Period Stable Matching in Cloud-Based Design and Manufacturing. In ASME 2018 International Design Engineering Technical Conferences and Computers and Information in Engineering Conference (pp. V01AT02A024-V01AT02A024). American Society of Mechanical Engineers.

31. Thekinen, J., & Panchal, J. H. (2017). Resource allocation in cloud-based design and manufacturing: A mechanism design approach. Journal of Manufacturing Systems, 43, 327-338.

32. Vate, J. H. V. (1989). Linear programming brings marital bliss. *Operations Research Letters*, *8*(3), 147-153.

33. Vohra, R. V. (2012). Stable matchings and linear programming. Current Science, 1051-1055.

34. Wang, X., Agatz, N., & Erera, A. (2018). Stable matching for dynamic ride-sharing systems. Transportation Science, 52(4), 850-867.

35. Zhang, R., Cheng, X., & Yang, L. (2017, September). Stable matching based cooperative v2v charging mechanism for electric vehicles. In 2017 IEEE 86th Vehicular Technology Conference (VTC-Fall) (pp. 1-5). IEEE.

# Appendices

## Appendix A

### Quantification of Contracts

Consider an order $d_i$ and a feasible supplier $s_j$ with $C_{ij}$ representing the set of contracts between them. Expected Utility theory is used for quantification of contracts in this problem setting. The following steps are required to apply the expected utility theory to this problem. A detailed description of the steps followed is provided in Fernandez et al. (2005). From an order's perspective, first the important attributes of a contract are selected, the utility function for each attribute is assessed and the individual utility functions are then combined using a multi attribute utility function. For instance, order $d_i$ values $p$ attributes $a = \{a_1, a_2 \ldots \ldots a_p\}$ in its contracts and the corresponding utility functions of these attributes are defined by $u_{i1}(a_1), u_{i2}(a_2) \ldots \ldots u_{ip}(a_p)$. The multi utility attribute function is then defined by equation (10).

$$u_i(a) = f[u_{i1}(a_1), u_{i2}(a_2) \ldots \ldots u_{ip}(a_p)] \tag{10}$$

Assuming order's preferences for the attributes to be independent, an additive multi attribute utility function can be used resulting in equation (11).

$$u_i(a) = \sum_{r=1}^{p} w_{ir} * u_{ir}(a_r) \tag{11}$$

where $w_{ir}$ is the scaling factor associated with attribute $a_r$ for order $d_i$. Similarly, considering suppliers' attributes and individual utility functions, the multi attribute utility function for a supplier can be defined as in equation (12). where $w_{jr}$ is a scaling factor associated with attribute $b_r$ for supplier $s_j$ which values $q$ attributes with individual utility function for $r_{th}$ attribute denoted by $u_{jr}(b_r)$.

$$u_j(b) = \sum_{r=1}^{q} w_{jr} * u_{jr}(b_r) \tag{12}$$

The following represents an illustrative example of utility calculation for an order supplier pair $(d_1, s_2)$ over terms of contract $c$. $s_2$ operates on a large scale, has a rating of 3 and is located in Atlanta, GA. $d_1$ located in Raleigh, NC requires a 3D design manufactured with aluminum alloy at a resolution of 200 microns due in 4 periods. The individual utility function of $d_1$ for location is $u_{11}(a_1) = 0.595 * a_1^2 - 1.516 * a_1 + 0.925$ over a range ($50 \leq a_1 \leq 500$) miles, size is $u_{12}(a_2) = 1$ if $a_2 = 'large', 0.6$ if $a_2 = 'medium', 0.3$ if $a_2 = 'small'$, rating is $u_{13}(a_3) = -0.219 * a_3^2 + 1.225 * a_3 - 0.005$ over a range ($1 \leq a_3 \leq 5$) and price is $u_{14}(a_4) = 0.922 * a_4^2 - 1.962 * a_4 + 1.033$ over a range ($640 \leq a_4 \leq 880$). The multi attribute utility function for $d_1$ is $u_1(a) = 0.2 * u_{11}(a_1) + 0.1 * u_{12}(a_2) + 0.3 * u_{13}(a_3) + 0.4 * u_{14}(a_4)$. Given the attributes of the contract with supplier $s_2$ (distance between the location of supplier and order is 400 miles, price quote is 750\$ and the order prefers to work with a large scale supplier), the utility $u_{ijc}$ of $d_1$ over the contract is calculated as $u_{1_2c} = 0.418$. For $s_1$, the individual utility function over material requirement of $d_1$ is $u_{11}(b_1) = 1$ if $b_1 = 'aluminum', 0.7$ if $b_1 = 'titanium', 0.3$ if $b_1 = 'steel'$, urgency of $d_1$ is $u_{12}(b_2) = -0.240 * b_2^2 + 1.329 * b_2 - 0.048$ over a range ($1 \leq b_2 \leq 8$) and revenue from the contract with $d_1$ is $u_{13}(b_3) = -0.444 * b_3^2 + 1.401 * b_3 + 0.032$ over a range ($150 \leq b_3 \leq 1600$). The multi attribute utility function for $s_1$ is $u_1(b) = 0.2 * u_{11}(b_1) + 0.3 * u_{12}(b_2) + 0.5 * u_{13}(b_3)$ and the utility $u_{jic}$ of $s_1$ over the contract is calculated as $u_{2_1c} = 0.611$. Similarly supplier and order utilities are calculated for all feasible contracts.

# Appendix B

## Influence of Rate of Order Arrivals

In this section, the impact of rate of arrival of orders in the systems is studied. As the order arrival rate is increased, the suppliers receive larger set of opportunities to choose from and the system becomes more competitive for orders. This can be seen in the average percentile rank achieved by the suppliers and orders in the system (Table 4). As the order arrival rate increases, the average percentile rank becomes better for the suppliers and worse for the orders in both AS and MWAS solutions. The impact of stability slightly reduces as the order arrival rate increases demonstrating that as the system become more competitive or capacity constrained, the loss in utility due to enforcing stability increases. However, the AS matching and the MWAS matching perform quite closely at different arrival rates.

The impact of order arrival rate on the pairwise and group transient stability of the solutions was also evaluated. It is not presented here considering space restrictions. Since the number of blocking pairs and groups in the AS and MWAS solutions are few, the impact of order arrival rate is arbitrary. In the MW formulation, as the order arrival rate increases, the number of blocking pairs and groups increases. With increased number of orders in the system, participants have more opportunities to form blocking groups and pairs. The percentage of unmatched participants in blocking pairs and groups increases for orders and decreases for suppliers again owing to more number of orders for the same set of suppliers and their capacities in the system.

Table 4: Influence of order arrival rate ($\lambda$) on system performance; Averages over 5 instances of 15 periods each (95% CI)

|  | AS | | | MWAS | | | MW | | |
|---|---|---|---|---|---|---|---|---|---|
| $\lambda$ | **80** | **100** | **120** | **80** | **100** | **120** | **80** | **100** | **120** |
| Impact of stability | 0.952 (0.002) | 0.947 (0.007) | 0.937 (0.006) | 0.96 (0.006) | 0.954 (0.004) | 0.944 (0.005) | 1 | 1 | 1 |
| Avg. order utility | 0.699 (0.012) | 0.679 (0.003) | 0.664 (0.008) | 0.698 (0.014) | 0.681 (0.003) | 0.669 (0.006) | 0.726 (0.004) | 0.73 (0.003) | 0.737 (0.002) |
| Avg. supplier utility | 0.659 (0.012) | 0.681 (0.006) | 0.698 (0.004) | 0.657 (0.012) | 0.679 (0.006) | 0.694 (0.003) | 0.662 (0.01) | 0.676 (0.006) | 0.685 (0.004) |
| Avg. allocated orders | 0.675 (0.019) | 0.731 (0.025) | 0.752 (0.021) | 0.682 (0.019) | 0.737 (0.025) | 0.757 (0.022) | 0.694 (0.02) | 0.748 (0.026) | 0.769 (0.02) |
| Matched orders | 0.844 (0.036) | 0.731 (0.016) | 0.63 (0.029) | 0.853 (0.036) | 0.738 (0.015) | 0.635 (0.028) | 0.869 (0.04) | 0.748 (0.015) | 0.645 (0.027) |
| Matched suppliers | 1 | 1 | 1 | 1 | 1 | 1 | 1 | 1 | 1 |
| Avg. order rank | 0.234 (0.028) | 0.282 (0.012) | 0.317 (0.011) | 0.237 (0.031) | 0.279 (0.009) | 0.308 (0.012) | 0.223 (0.013) | 0.234 (0.006) | 0.236 (0.005) |
| Avg. supplier rank | 0.493 (0.029) | 0.413 (0.026) | 0.339 (0.006) | 0.497 (0.028) | 0.422 (0.026) | 0.356 (0.012) | 0.518 (0.016) | 0.465 (0.016) | 0.414 (0.008) |